\documentclass[Journal]{IEEEtran}
\usepackage[compress]{cite}
\usepackage{epsfig,graphics,mathrsfs,amssymb,amsmath,bm,color,yfonts,txfonts,multirow,slashbox}
\usepackage{subfigure}
\usepackage{algorithm2e}
\makeatletter
\renewcommand{\citepunct}{,\penalty\@m\hskip.13emplus.1emminus.1em}
\renewcommand{\citedash}{\hbox{--}\penalty\@m}
\makeatother

\begin{document}

\title{Feedback Bits Allocation for Interference Minimization in Cognitive Radio Communications }

\author{ Mirza Golam Kibria, Fang Yuan and Fumihide Kojima
 \thanks{M. G. Kibria and F. Kojima are with Wireless Network Research Institute, National Institute of Information and Communications Technology (NICT), Japan (E-mails: \{mirza.kibria, f-kojima\}@nict.go.jp).
F.~Yuan was with Wireless Network Research Institute, National
Institute of Information and Communications Technology (NICT), Japan
(E-mail: yuanfangv@163.com). } }

\maketitle

\begin{abstract}
This letter studies the limited feedback cognitive radio system,
where the primary users (PU) are interfered by the secondary
transmitter (ST) due to the imperfect beamforming. We propose to
allocate the feedback bits among multiple PUs to minimize the
maximum interference caused by the ST, by exploiting the heterogeneous
average channel gains. In addition, we study the problem of
minimizing the total feedback bits under a predefined interference
threshold at the PUs. The solutions with low complexity are proposed
for the studied problems, and the performances of bit allocations
are analyzed. Simulation results validate our analysis and
demonstrate that the proposed solutions work very well in terms of
minimizing the maximum interference caused by the ST and minimizing
the total feedback bits under predefined interference threshold at
the PUs for limited feedback CR system.
\end{abstract}
%=================================================================================
%=================================================================================
\section{Introduction}
%=================================================================================
%=================================================================================
\IEEEPARstart{C}{ognitive} radio (CR) system is known as one of the
promising techniques to meet the ever-increasing demand on spectrum
efficiency (SE) in the future wireless systems
\cite{Haykin,Villardi}. Depending on spectrum sharing strategies,
there are generally two operation modes in CR systems. One is
overlay mode, where the transmission in secondary system is enabled
only when primary system is not on transmission \cite{Haykin}. The
other is underlay mode, where the transmission in secondary system
is enabled if the interference to primary system can be tolerated
\cite{Zhang2010}. Compared to the overlay mode, the underlay mode
has more potential in improving the SE as it allows more chances for
simultaneous cognitive transmissions \cite{Zhang2010}. Thus this
work focuses on the underlay mode for CR system.

Among all, the CR system with multiple antennas at the secondary
transmitter (ST) is capable of applying the beamforming at the ST to
improve the secondary transmission while protecting the primary
users (PU) \cite{Taherpour2010}. In these CR systems, the channel
direction information from the ST to the PU (CDIsp) must be acquired
at the ST to assist the beamforming, and the acquisition of CDIsp
results in additional signaling overhead between the primary and
secondary systems.

To reduce the signaling overhead for the CDIsp in CR system, limited
feedback techniques have been proposed, where the CDIsp is quantized
at the PU and fed back to the ST \cite{Huang2011,Chen2014}. The
obtained CDIsp at the ST is imperfect because of the quantization,
and thus the interference is residual to the PU that can not be
completely nulled after the beamforming. Theoretically, allocating
more feedback bits for each PU can provide more accurate CDIsp and
reduce the residual interference, which is however constrained by
the finite feedback capacity between primary and secondary systems
\cite{Chen2014}. To avoid the severe interference for the PU, it is
important to minimize the maximum interference in primary system
caused by the ST.

In this letter, we study the problem of allocating the feedback bits
among multiple PUs to minimize the maximum interference caused by
the ST in limited feedback CR system. We also minimize the total
feedback bits budget under a predefined interference threshold at
the PUs. For both two problems, the solutions are provided in
closed-form and of low complexity. Notation: we use $|~|$, $()^T$,
$E\{\}$ to denote the absolute value, the Hermitian operator and the
expectation operator, respectively.

\section{System Model}
Consider the ST equipped with a number of $N$ antennas serves a
secondary user (SU) by beamforming while mitigating the
interferences to a number of $K$ PUs in the primary system. The
received instantaneous interference at the PU $k$ from the ST is
expressed as
\begin{align}
I_{k}&=|\sqrt{\lambda_k}\pmb{h}_{k}^H\pmb{v}_{0}d_{0}|^2,\nonumber
\end{align}
where $k=1,\ldots,K$, $\lambda_k$ and $\pmb{h}_{k}$ corresponds
respectively to the average channel gain and instantaneous channel
vector from the ST to PU $k$, $\pmb{v}_{0}$ and $d_{0}$ are
respectively the unit-norm beamformer and data symbol at the ST for
the SU. We assume the channel vectors $\pmb{h}_{k}$ are subject to
identically independent distributed (i.i.d.) flat Rayleigh fading,
and the PUs have perfect channel information about $\pmb{h}_{k}$
after the channel estimation.

Denote the perfect CDIsp as
$\bar{\pmb{h}}_{k}=\pmb{h}_{k}/|\pmb{h}_k|$, which is unit norm and
conveys only channel direction information \cite{Jindal2006}.  In
the limited feedback literature \cite{Jindal2006,Yoo2007}, the CDIsp
is firstly quantized through a given codebook with a proper size at
each PU, and then the bits for the quantized CDIsp are fed back to
the ST for beamforming. There are many protocols to support
forwarding the CDIsp from the primary system to the secondary
system, e.g., the S1 protocol in \cite{Sref} when primary and
secondary systems are deployed in macro and pico cells, respectively.

Let $\hat{\pmb{h}}_{k}$ be the quantized version of
$\bar{\pmb{h}}_{k}$. The relation between perfect CDIsp and
quantized CDIsp is given as \cite{Yoo2007}
\begin{align}
\bar{\pmb{h}}_{k} =
\cos{\theta}_{k}\hat{\pmb{h}}_{k}+\sin{\theta_{k}}\pmb{q}_{k},\nonumber
\end{align}
where $\pmb{q}_{k}$ is the quantization error vector, and
$\cos^2{\theta_{k}}=|\bar{\pmb{h}}_{k}^H\hat{\pmb{h}}_{k}|^2$
reflects the accuracy of CDIsp received at the ST.

The average CDIsp distortion due to quantization is defined as
\cite{Jindal2006}
\begin{align}
{\delta_k}&=1-{\rm{E}}\{|\bar{\pmb{h}}_{k}^H\hat{\pmb{h}}_{k}|^2\}={\rm{E}}\{\sin^2{\theta_{k}}\}.\nonumber
\end{align}
The value of $\delta_k$ depends on the employed codebooks. For the
tractability, we apply the quantization upper bound method
introduced in \cite{Yoo2007}, which assumes each quantization cell
is a Voronoi region of a small spherical cap. From Eq. (13) of
\cite{Yoo2007}, we find the average quantization error as
\begin{align}
{\delta_k}=\frac{N-1}{N}2^{-\frac{b_k}{N-1}},\label{aql}
\end{align}
where  $b_k$ is the number of bits allocated to user $k$ for
quantizing its CDIsp $\bar{\pmb{h}}_{k}$. The work in \cite{Yoo2007}
demonstrates via simulation that \eqref{aql} provides a good
approximation for any other codebooks designed for i.i.d. flat
Rayleigh channels.

As in most limited feedback literature \cite{Jindal2006,Yoo2007}, we
consider the zeroforcing (ZF) beamforming is applied at the ST to
reduce the interference into the primary system, i.e., the
beamforming vector is selected such that
$|\pmb{h}_k^H\pmb{v}_0|^2=0$ for all $k$ under perfect CDIsp.
Denoting $\pmb{h}_0$ as the channel vector from the ST to the SU,
one well-known ZF beamforming scheme is from the pseudo-inverse by
normalizing the first column of the matrix
$\pmb{V}=\pmb{H}[\pmb{H}^H\pmb{H}]^{-1}$ where $\pmb{H}=[\pmb{h}_0,
\hat{\pmb{h}}_1, \cdots, \hat{\pmb{h}}_K]$. Under limited feedback,
although the interference cannot be nulled completely due to
imperfect CDIsp, the ZF beamforming has the merit of simplicity and
robustness to imperfect CDIsp. In ZF scheme, when the elements in
$\pmb{h}_k$ are i.i. d. with a unit variance, it is shown in
\cite{Yuan2013},
\begin{align}
E\{|\pmb{h}_k^H\pmb{v}_0|^2\}=\delta_k.\nonumber
\end{align}
Then the average interference from the ST to the PU $k$ under
limited feedback becomes
\begin{align}
I_{av,k}=E\{I_k\}=P_0\lambda_k\delta_k,\label{outs}
\end{align}
where the transmit power at the ST is $E\{|d_0|^2\}=P_0$.

\section{Problems and Solutions}

\subsection{Interference Minimization Under Limited Feedback}
Observing \eqref{outs}, allocating more feedback bits to the PU $k$
reduces its residual interference from the ST. Yet the feedback
capacity from the primary system to secondary system is usually
finite and shared among the users. The interferences in \eqref{outs}
are linear with average channel gains $\lambda_k$, which are
heterogenous for individual PUs. To avoid the severe interference to
primary system, it is necessary to allocate the total available
feedback bits among multiple PUs by considering the heterogeneity of
average channel gains $\lambda_k$.

Motivated by this, the problem of allocating feedback bits among
multiple PUs to minimize the maximum interference in limited
feedback CR system under the sum-bit constraint can be
described as
\begin{align}\label{optimalPM}
\min_{b_k} &\quad \max_{k\in\{1,\ldots,K\}} I_{av,k}\\
\text{s.t.}  & \quad \sum\nolimits_{k=1}^Kb_k\leq B,\label{ccc1}\\
 & \quad b_k\geq 0, \quad \forall k,\label{ccc2}
\end{align}
where $B$ is the total number of feedback bits for all $K$ PUs in
the available feedback capacity from the primary system to the
secondary system.

The problem in \eqref{optimalPM} is convex after relaxing the
integers $b_k$ into the continuous variables, since the second-order
derivative of the objective function $I_{av,k}$ with regard to $b_k$
is positive and the constraints in \eqref{ccc1} and \eqref{ccc2} are
linear. Then, the problem in \eqref{optimalPM} can be solved via
standard convex optimization tools \cite{Boyd2004}. After obtaining
the optimal solution to \eqref{optimalPM2}, the allocation results
can be rounded into nearest integers as the number of bits allocated
for each PU \cite{Guo2002}.

To provide an explicit solution with low complexity for the bit
allocation in \eqref{optimalPM}, we further consider a suboptimal
problem. Inspired by the equation
\begin{align}
\lim_{L\rightarrow\infty}\left(\sum_{k=1}^KI_{av,k}^L\right)^{\frac{1}{L}}=\max_{k\in\{1,\ldots,K\}}
I_{av,k},\label{ident}
\end{align}
we try to optimize the new objective
$\left(\sum_{k=1}^KI_{av,k}^L\right)^{\frac{1}{L}}$ instead of
$\max_{k\in\{1,\ldots,K\}}I_{av,k}$ under a sufficiently large
positive integer $L$.

Then the new optimization problem becomes
\begin{align}\label{optimalPM2}
\min_{b_k} &\quad \left(\sum_{k=1}^KI_{av,k}^L\right)^{\frac{1}{L}}\\
\text{s.t.}  & \quad \sum\nolimits_{k=1}^Kb_k\leq B,\\
 & \quad b_k\geq 0, \quad \forall k.
\end{align}
The problem in \eqref{optimalPM2} is identical to that in
\eqref{optimalPM} when $L$ approaches infinity as guaranteed by
\eqref{ident}. However, under a finite $L$, the problem in
\eqref{optimalPM2} is suboptimal to that in \eqref{optimalPM} since
$\left(\sum_{k=1}^KI_{av,k}^L\right)^{\frac{1}{L}}\leq
\max_{k\in\{1,\ldots,K\}}I_{av,k}$ always holds.

The problem in \eqref{optimalPM2} is still convex, since $L$ is a
constant and the inside term $I_{av,k}$ is convex on $b_k$. The
Lagrangian function of the problem in \eqref{optimalPM2} is
\begin{align}
L(b_k,\nu_0)=\left(\sum\nolimits_{k=1}^KI_{av,k}^L\right)^{\frac{1}{L}}+\nu_0\left(\sum\nolimits_{k=1}^Kb_k-
B\right),
\end{align}
where $\nu_0$ is the lagrangian multiplier. Thus, the optimal
solution to the problem \eqref{optimalPM2} should satisfy the
Karush-Kuhun-Tucker (KKT) conditions
\begin{align}
\frac{\partial L(b_k,\nu_0)}{\partial
b_k}&=-c\lambda_k^L2^{-\frac{Lb_k}{N-1}}+\nu_0=0,\quad \forall
k\\\frac{\partial L(b_k,\nu_0)}{\partial
\nu_0}&=\sum\nolimits_{k=1}^Kb_k-B=0,
\end{align}
where
$c=\ln2\frac{(N-1)^{L-1}P_0^L}{N^L}\left(\sum\nolimits_{k=1}^KI_{av,k}^L\right)^{\frac{1}{L}-1}$.
By introducing a new positive parameter $\nu=\frac{L\nu_0}{c(N-1)}$,
the KKT conditions can be written more concisely as
\begin{align}
b_{k} = \frac{(N\!-\!1)}{L}\left[
L\log_2\lambda_{k}-\!\log_2\frac{\nu(N\!-\!1)}{L}\right]^\dagger,\label{BAR}
\end{align}
where $\nu>0$ should satisfy $\sum_{k=1}^{K}b_{k}= B$, and
$[x]^\dagger=\max\{0,x\}$. From \eqref{BAR}, it is clear that the
optimal solution to the problem \eqref{optimalPM2} allocates more
bits to quantize the CDIsp from the PU who has a larger average
channel gain $\lambda_{k}$.

It is known that the solution satisfying \eqref{BAR} can be found by
the standard water-filling algorithm \cite{Palomar2005}, which can
be implemented with one-dimensional search over $\nu$ and converges
very fast with less than $K$ iterations. After $\nu$ is found, the
bit allocation results can be immediately obtained from \eqref{BAR},
which only need regular scalar operations. Therefore, the
computational complexity of the solution to \eqref{optimalPM2} is
reduced from the solution of standard optimization tools to
\eqref{optimalPM}. The solution to \eqref{optimalPM} can serve as
the performance upper bound for any bit allocation to minimize the
maximum interference in limited feedback CR system. In practice, the
suboptimal solution in \eqref{BAR} may be more desirable for CR
system which is sensitive to the computational complexity. Note that
although the optimal setting $L$ should be infinity, it is
sufficient to set a finite large $L$ in the optimization for
achieving a good result in practice as shown in the simulation
section.

The suboptimal problem in \eqref{optimalPM2} allows us to obtain the
relationship between the minimized maximum interference and the
number of total feedback bits under asymptotical analysis. To see
this, we consider that the total number feedback bits $B$ is large
such that the operation $[]^\dagger$ can be removed in \eqref{BAR}.
In this case, it can be verified that the solution in \eqref{BAR}
achieves the same optimization result as the algorithmic inequality
$\sum_{i=1}^Kx_i/K\geq(\prod_{i=1}^Kx_i)^{1/K}$ given by
\begin{align}
\left(\sum_{k=1}^KI_{av,k}^L\right)^{\frac{1}{L}}&\geq
K^{\frac{1}{L}}\left(\prod_{k=1}^KI_{av,k}\right)^{\frac{1}{K}}=K^{\frac{1}{L}}P_0\left(\prod_{k=1}^K\lambda_k\delta_k\right)^{\frac{1}{K}}\nonumber\\
&=K^{\frac{1}{L}}\frac{N-1}{N}P_0\left(\prod_{k=1}^K\lambda_k\right)^{\frac{1}{K}}2^{-\frac{B}{K(N-1)}},
\end{align}
where the last equality is led by
$\left(\prod_{k=1}^K\delta_k\right)^{\frac{1}{K}}=\frac{N-1}{N}2^{-\frac{B}{K(N-1)}}$
under the sum-bit constraint that $\sum_{k=1}^{K}b_{k}= B$.

Since the problem in \eqref{optimalPM} and \eqref{optimalPM2} are
identical when $L$ approaches infinity, by using
$\lim_{L\rightarrow\infty}K^{\frac{1}{L}}=1$, the optimized maximum
interference in \eqref{optimalPM} under a large total number of
feedback bits $B$ satisfies
\begin{align}
I_{opt}=\lim_{L\rightarrow\infty}\left(\sum_{k=1}^KI_{av,k}^L\right)^{\frac{1}{L}}=\frac{N-1}{N}P_0\left(\prod_{k=1}^K\lambda_k\right)^{\frac{1}{K}}2^{-\frac{B}{K(N-1)}},\label{opti}
\end{align}
which is linear with the geometric mean of the average channel gains
among the PUs, and degrades exponentially as the number of antenna
$N$ increases.

\subsection{Feedback Minimization Under Interference Threshold}
Another challenge in the design for CR systems is to guarantee that
interferences at the PUs caused by the ST is restricted to below a
predefined interference threshold, $I_{\rm{max}}$.

In such systems, it is more desirable to minimize the total required
feedback bits such that the interference to each PU is below a given
threshold. The corresponding feedback bits budget optimization
problem can be modeled as
\begin{align}\label{optimalPMX}
\min_{b_k} &\quad \sum\nolimits_{k=1}^Kb_k\\
\text{s.t.}  & \quad I_{av,k}\leq I_{{\rm{max}}}.\label{cccN1}\\
 & \quad b_k\geq 0, \quad \forall k.\label{cccN2}
\end{align}

The above problem can be solved easily. To satisfy the constraints
in \eqref{cccN1} and \eqref{cccN2}, according to \eqref{aql} and
\eqref{outs}, it requires that
\begin{align}
b_k\geq
(N-1)\left[\log_2{P_0\lambda_k}-\log_2\left(\frac{NI_{{\rm{max}}}}{N-1}\right)\right]^\dagger.\label{sbopt1}
\end{align}
The optimal solution to \eqref{optimalPMX} is to set $b_k$ as the
smallest integer satisfying \eqref{sbopt1} for each PU, and the
minimum required total feedback bits obtained under the unrounded
$b_k$ is
\begin{align}
\sum_{k=1}^Kb_k=\sum_{k=1}^K
(N-1)\left[\log_2{P_0\lambda_k}-\log_2\left(\frac{NI_{{\rm{max}}}}{N-1}\right)\right]^\dagger,\label{sbopt}
\end{align}
which means the number of required total feedback bits increases
with the average channel gain $\lambda_k$ but decreases with the
interference threshold $I_{{\rm{max}}}$.

\section{Simulation Results}
The proposed bit allocation solutions are evaluated with
simulations. We consider the case of three PUs in the primary
system, i.e., $K=3$. The average channel gains from the ST to  three
PUs are respectively $\lambda_1,\lambda_2,\lambda_3$, and the
transmit power at the ST is set as $P_0=1$. The codebooks for
quantizing the CDIs are adopted as given in
\cite{Jindal2006,Yoo2007}.

\begin{figure}
   \centering
   \includegraphics[scale=.53]{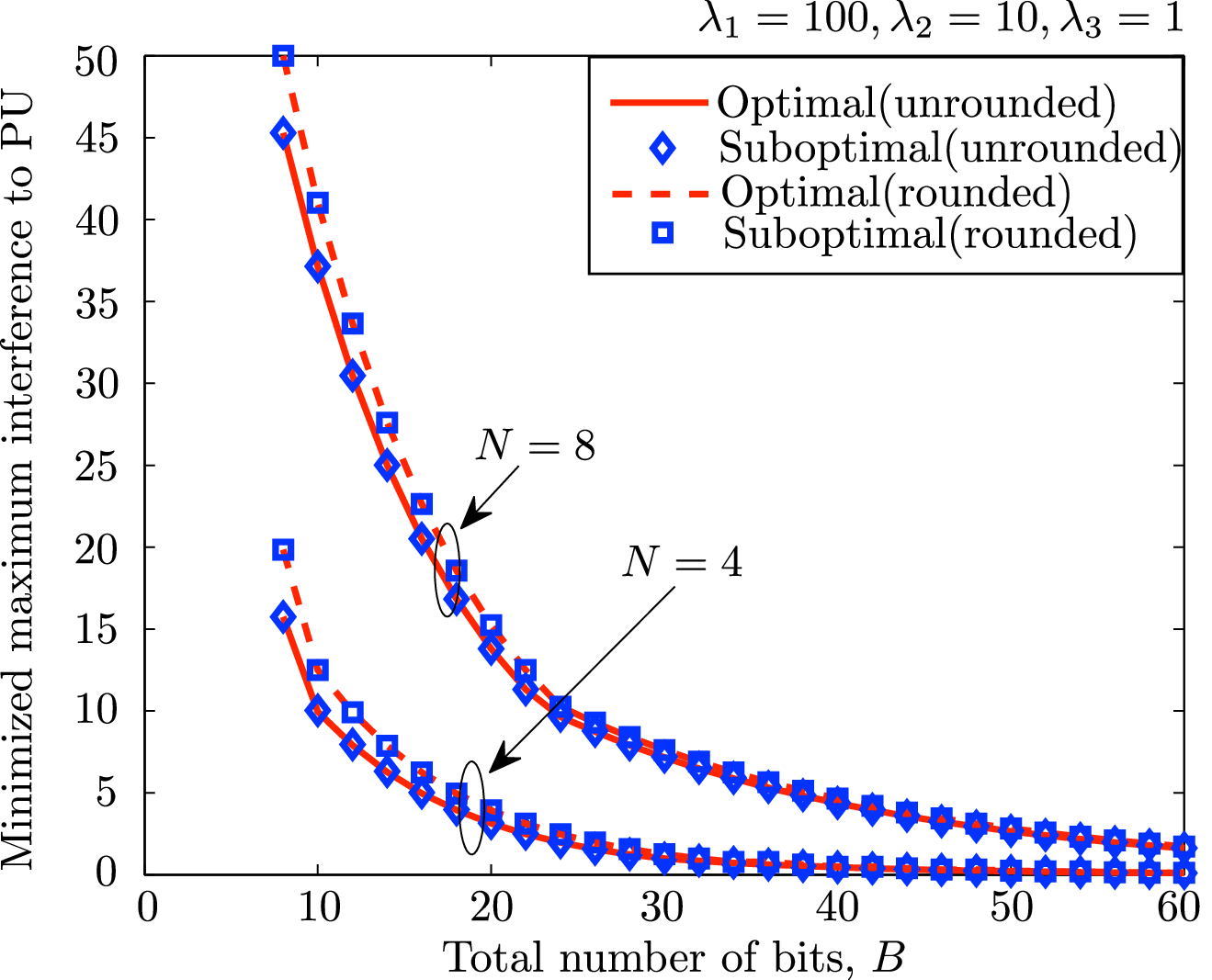}
   \caption{The minimized maximum interference versus different value of $B$ under $N=4,8$.}\label{prd1}
\end{figure}

\begin{figure}
   \centering
   \includegraphics[scale=.54]{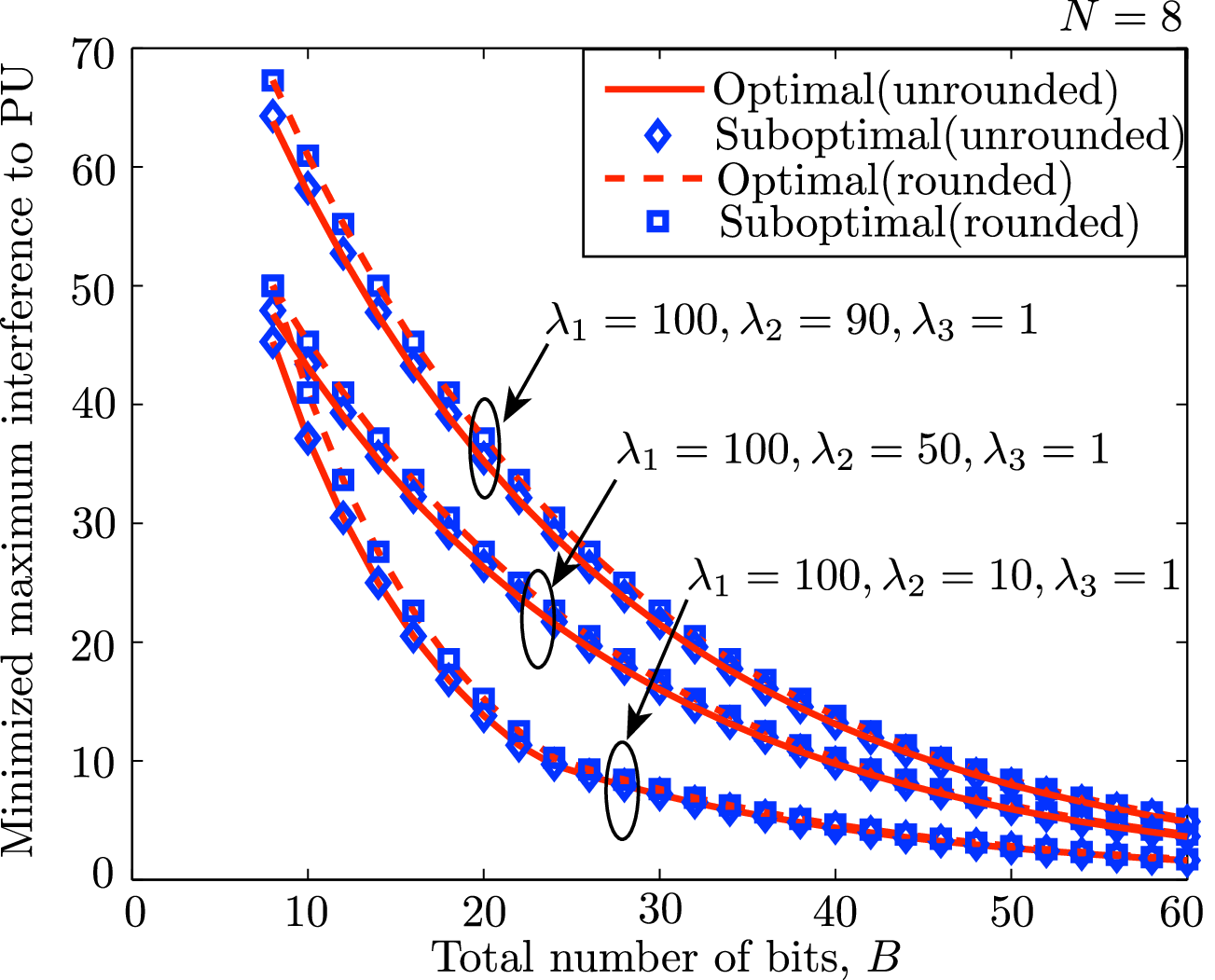}
   \caption{The minimized maximum interference versus different value of $B$ under different average channel gain.}\label{prd2}
\end{figure}

\begin{figure}
   \centering
   \includegraphics[scale=.54]{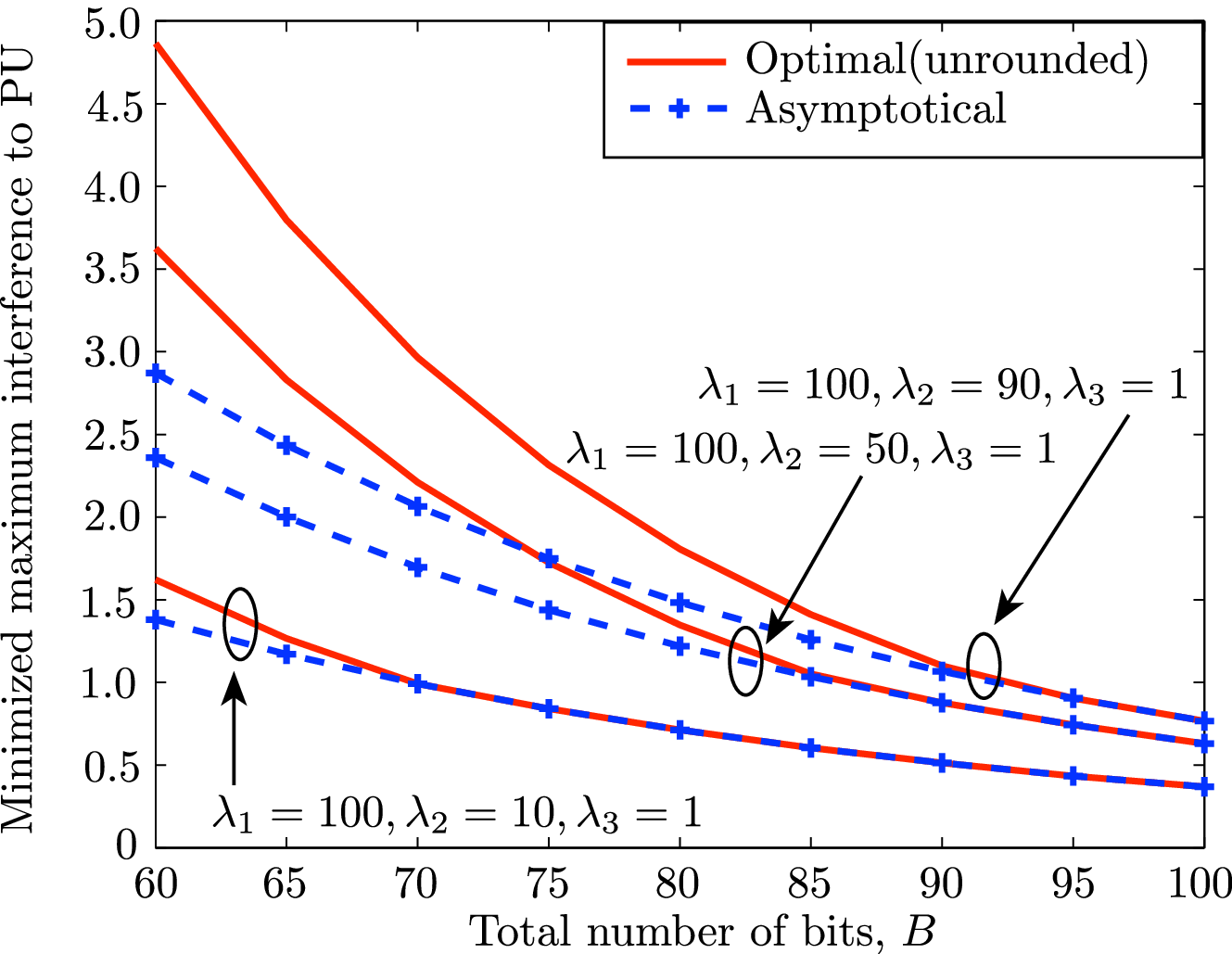}
   \caption{Comparison of the minimized maximum interference under the optimal solution and the asymptotical analysis.}\label{prd5}
\end{figure}

The minimized maximum interferences versus different value of $B$
under $N=4,8$ are provided in Fig.~\ref{prd1}, where
$\lambda_1=100$, $\lambda_2=10$ and $\lambda_3=1$. The optimal
solution in \eqref{optimalPM} and suboptimal solution in
\eqref{optimalPM2} by setting $L=100$ are investigated, and both the
rounded and unrounded results are provided. As shown in the figure,
we can find that the proposed suboptimal solution has almost the
same performances as the optimal one for both rounded and unrounded
cases. The minimized maximum interference drops to zero as the total
number of feedback bits increases, while the descending rate is
dominated by the number of antennas $N$ at the ST as indicated by
the result in \eqref{opti}.

The minimized maximum interferences versus different values of $B$
under different average channel gain are provided in Fig.
\ref{prd2}, where $\lambda_1=100$ and $\lambda_3=1$ are fixed,
$\lambda_2$ varies in three cases, i.e., $\lambda_2=90$,
$\lambda_2=50$ and $\lambda_2=10$. As shown in the figure, we can
find that the minimized maximum interference increases as the
average channel gain $\lambda_2$ increases.

The minimized maximum interferences under the optimal solution to
\eqref{optimalPM} and the asymptotical analysis in \eqref{opti} are
further compared in Fig. \ref{prd5}, where the setting is similar to
Fig.~\ref{prd2} but with larger number of total feedback bits. As
shown in the figure, we can find as the total number of bits $B$
gets larger, the minimized maximum interference converges to the
asymptotical analysis in \eqref{opti} gradually under different
average channel gains, which validates our analysis in \eqref{opti}.

\begin{figure}
   \centering
   \includegraphics[scale=.55]{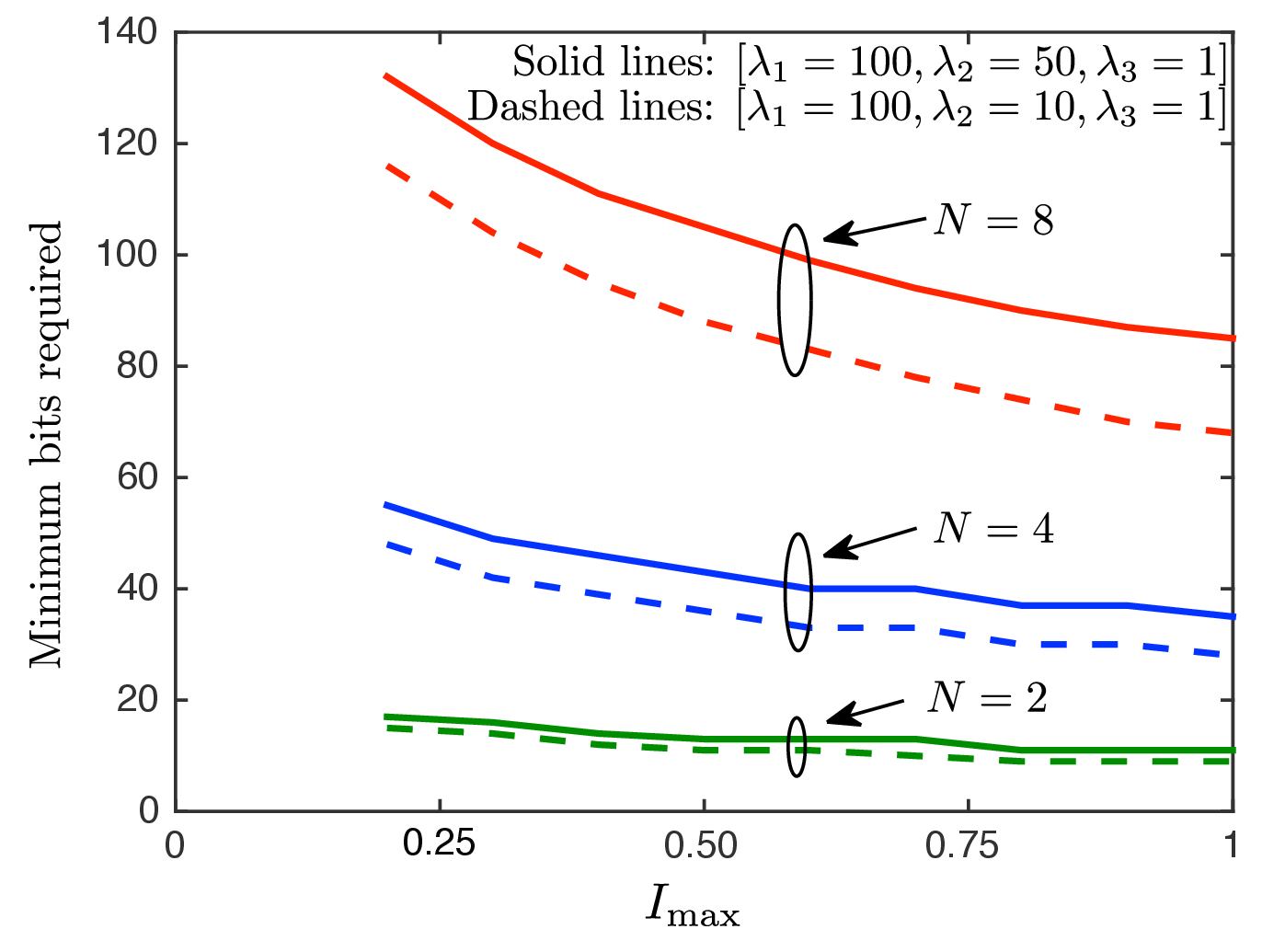}
   \caption{Minimum feedback bits requirement versus varying interference threshold $I_{\rm{max}}$.}\label{prd3}
\end{figure}

Finally, we evaluate the performance of the proposed solution for
the total feedback bits minimization under the interference
threshold in Fig.~\ref{prd3}, with $N=2,4,8$ respectively. As shown
in the figure, as the interference threshold becomes larger, the
minimum required number of total feedback bits decreases. Moreover,
the minimum required number of total feedback bits increases with
the average channel gains under different number of antennas as we
have analyzed.

\section{Conclusions}
We have studied the problem of allocating the feedback bits for
minimizing the maximum interference in CR system. The solution is
proposed with low complexity and analyzed in asymptotic regime. It
has revealed the minimized maximum interference is linear with the
geometric mean of the average channel gains, and drops to zero
exponentially as the number of feedback bits increases, while the
descending rate is dominated by the number of antennas at the ST. We
have also studied the total feedback bits minimization problem under
a predefined interference threshold at the PUs, where the minimum
required number of total feedback bits decreases as the interference
threshold becomes larger. Simulation results demonstrate that the
proposed schemes work very well for CR systems.

\end{document}